\documentclass[aps,letterpaper,prl,footinbib,floatfix,showpacs,twocolumn]{revtex4}
\usepackage{amsmath}
\usepackage{amsfonts}
\usepackage{amssymb}
\usepackage[scanall]{psfrag}
\usepackage{psfrag}
\usepackage{graphicx}
\usepackage{color}
\usepackage{subfigure}

\newcommand{\p}[1]{\left( {#1} \right)}
\newcommand{\br}[1]{\left[ {#1} \right]}
\newcommand{\avg}[1]{\left<{#1}\right>}
\renewcommand{\vec}[1]{\boldsymbol{#1}}

\newcommand{\unit}[1]{\ensuremath{\, \mathrm{#1}}}

\newcommand{\dg}{^{\dagger}}

\newcommand{\gb}{\beta}
\newcommand{\gd}{\delta}

\newcommand{\gl}{\lambda}

\newcommand{\gs}{\sigma}

\newcommand{\bra}[1]{\left\langle #1 \right\rvert}
\newcommand{\ket}[1]{\left\lvert #1 \right\rangle}
\newcommand{\braket}[2]{\left\langle #1 \mid #2 \right\rangle}

\newcommand{\up}{\uparrow}
\newcommand{\dn}{\downarrow}

\DeclareMathOperator{\Tr}{Tr}

\begin{document}

\author{Yariv Yanay}
\affiliation{Laboratory of Atomic and Solid State Physics, Cornell University, Ithaca NY 14850}
\author{Erich J. Mueller}
\affiliation{Laboratory of Atomic and Solid State Physics, Cornell University, Ithaca NY 14850}
\author{Veit Elser}
\affiliation{Laboratory of Atomic and Solid State Physics, Cornell University, Ithaca NY 14850}
\title{Magnetic polarons in two-component hard core  bosons}
\date{\today}

\begin{abstract}
We use a high-temperature expansion to explore spin correlations around a single hole in a two-dimensional lattice filled with a hard-core two component bose gas. We find that the spins around the hole develop ferromagnetic order and quantify the degree of polarization at temperatures of order the hopping energy, finding a measurably nonzero polarization. We also discuss the effect of fixing the overall magnetization of the system for finite-sized systems.
\end{abstract}


\maketitle

In the mid 1960's, Thouless and Nagaoka studied the two-component Fermi system on a bipartite lattice with very strong on-site repulsion \cite{Thouless1965,Nagaoka1966}.  They found that in the presence of a single hole the ground state was a fully polarized ferromagnet.  
These and further studies showed that at finite temperatures the system is not fully polarized: near the hole there is a ferromagnetic ``bubble", while far away the spins are uncorrelated \cite{Heritier1977,Andreev1976}.
On such bipartite lattices, the statistics are irrelevant for the single-hole problem, and the same physics should be seen in the bosonic case as in the fermionic system. 
Thus the bosonic ground state is the Nagaoka state, and at finite temperature, ferromagnetic correlations are found near the hole.
Here we calculate these correlations in a two-component gas of hard core lattice bosons.  We find that at experimentally relevant temperatures these correlations are measurable using a quantum gas microscope \cite{Bakr2009}.

This is the simplest example of emergent physics in a strongly correlated system.  Variants of it are also highly nontrivial: for example the ground state of two component fermions on a non-bipartite lattice with a single hole is unknown. 
A qualitative  picture of this ferromagnetism can be developed by imagining a child's puzzle where tiles slide on a square grid.  One tile is missing.  By moving this ``hole" one can rearrange the tiles.  Here we have a quantum mechanical version of this puzzle.  The motion of the hole from one location to another involves summing all possible paths.  If the tiles are in a symmetric superposition of all possible arrangements (corresponding to ferromagnetism) then these paths will add constructively, allowing the hole to move over large distances.  This ferromagnetic arrangement thereby minimizes the zero-point energy of the hole.

Borrowing the term from how electronic motion couples to lattice distortions, the elementary excitation consisting of a hole dressed by a ferromagnetic cloud is referred to as a ``polaron''.
Other cold-atom polaron problems include the behavior of a single down-spin atom in a Fermi sea of up-spins \cite{Sommer2011,Prokofev2008,Lobo2006,Pilati2010}.

Even far from the strong-coupling hard-core limit studied here the physics of two-component bosons is quite rich.  
This physics has been explored in theoretical works \cite{Kaurov2005,Kuklov2004,Kuklov2003,Fil2004,Chung2012}, and in  
 cold gas experiments \cite{Gadway2010,Catani2008}.  The components can be different hyperfine states \cite{Gadway2010}, or different atomic species \cite{Catani2008}.
In the most ordered state there will be two independent order parameters, and it costs energy to twist the phase $\phi_1,\phi_2$, of either condensate.  Depending on interaction parameters one can also find states where only some linear combination of the two phases has a finite stiffness.  For example, with sufficiently strong attraction between the species there will be a condensate of ``pairs" but no single particle condensate \cite{Nozieres1982}: One then has a stiffness to twisting $\phi_1+\phi_2$, but not $\phi_1-\phi_2$.  Even more exotic is the ``counter-superfluid'' phase formed when the interspecies repulsion becomes strong: One then has a stiffness to twisting $\phi_1-\phi_2$.   Under these circumstances trying to drive a current of species $1$ to the right creates a current of species $2$ to the left.  
Identifying the two components as the $\pm z$-component of a  pseudospin-1/2 object -- the counter-superfluid state corresponds to an $x-y$ ferromagnet.
If the in-species interactions are not sufficiently strong, either of these exotic states can be preempted by phase separation or collapse. \cite{Ozaki2012}
In the single hole limit, the phase stiffnesses scale as the inverse of the system size.

Here we use a high temperature expansion to calculate the correlations between spins bordering a single hole in a two-component hard-core Bose system on the square lattice.  Using the techniques in \cite{Itah2009,Gericke2008,Karski2009,Nelson2007,Bakr2009, Fukuhara2012}, these correlations can be directly measured, giving a signature of this interesting physics. 

The temperature scale at which these correlations become significant is of order the hopping  energy $t$. In physical units, this energy is on the order of $t \sim k_{B}\times \p{1 \unit{nK}}$ for $^{87}$Rb atoms trapped by $\gl = 820 \unit{nm}$ lasers \cite{Spielman2007}, but using lighter atoms such as $^{7}$Li would increase the hopping energy energy and corresponding temperature by a factor of ten. Similarly, using a shorter wavelength lattice would also increase this scale. 

Our study assumes hard-core interactions, where double occupancy is forbidden. In most experiments, the strength of on-site interactions $U$ is fixed and the hard-core regime is achieved by increasing the height of the potential barrier between neighboring sites so that $t\ll U$. Corrections to the hard-core results scale as $t/U$. Spielman et al. \cite{Spielman2007} report results with $t/U\sim 0.001$.

Another relevant experimental detail is most cold atom systems are confined in harmonic traps. Local physics, such as the correlations we study, are unaffected by such confinement, as long as one restricts attention to regions where the polarons are dilute.

The physics of Nagaoka ferromagnetism is relevant for a number of other cold atom systems \cite{VonStecher2010,Okumura2011}.

\section{Analysis}
We model the two-component Bose system via the single band Bose-Hubbard Hamiltonian
\begin{equation} \begin{split}
\hat H = -t\sum_{\avg{i,j}}\sum_{\gs=\up,\dn} \p{a_{\gs,i}\dg a_{\gs, j} + a_{\gs,j}\dg a_{\gs, i}}
\label{eq:BH}
\end{split} \end{equation}
where $a_{\gs,i}$ ($a_{\gs,i}\dg$) is the bosonic annihilation (creation) operator for a particles of type (``spin'') $\gs$ at lattice site $i$, and $\avg{i,j}$ are all nearest-neighbor pairs and we limit ourselves to a two-dimensional square lattice. The single-particle spectrum has band-width of $8 t$.  
We work in the canonical ensemble, with fixed particle number, and do not need to include a chemical potential.

The Bose-Hubbard model is a good description of the system as long as the band-spacing $E_{b}$ is large compared with the other relevant energy scales. We require $t\ll E_{b}$ so that the (single-particle) bands are distinct, while $T\ll E_{b}$ is required so that all bosons are in the lowest band. In addition, we will be analyzing Eq.~(\ref{eq:BH}) within a high temperature expansion, requiring that the ratio $T/t$ is not too small.

For a cold-atoms experiment described by the single-band Hubbard model, the band spacing varies with  microscopic parameters as $E_{b}\sim \sqrt{V_{0} E_{R}}$, where $V_0$ is the height of the potential barriers between lattice sites, $E_{R} = \hbar^{2}k^{2}/2m$: $k=2\pi/\gl$ being the laser wavenumber and $m$ the particle mass.  The tunneling $t$ depends exponentially on $V_0$ and is typically $t\sim 0.1-0.01 E_{R}$ for $V_0\gtrsim E_R$ \cite{Jaksch1998}.   There is therefore a separation of scales, allowing $t\sim T\ll E_b$. Deeper lattices accentuate this separation, at the cost of requiring lower temperatures.

Strong interactions imply a hard-core constraint
\begin{equation} \begin{split}
a_{\gs,i}\dg a_{\tau,i}\dg = 0,
\end{split} \end{equation}
which is valid when the on-site interactions are large compared to $t$.

We examine the case of a single hole in an infinite system and calculate the finite temperatures expectation values of an 
observable operator $\hat X$ by
\begin{equation} 
\avg{X}  = \frac{1}{Z}\Tr \hat X e^{-\gb \hat H};\quad
 Z  = \Tr e^{-\gb \hat H}
 \end{equation}
where $\gb = \p{k_{B}T}^{-1}$ is the inverse temperature; we take $k_{B}=1$. 
The trace is most readily calculated in a basis given by placing the hole on the site $r_h$, and specifying the pseudospin $\sigma_{i}=\uparrow/\downarrow$ on all remaining sites $i\neq r_h$.  We will look at the correlations between spins on positions which are fixed relative to $r_h$.  For observables of that form, denoting by $\xi$ a spin state with the hole at the origin, we have
\begin{equation} \begin{split}
\avg{X} & = \frac{N_{sites}}{Z}\sum_{\xi} X\p{\xi} \bra{\xi} e^{-\gb \hat H}\ket{\xi}
\end{split} \end{equation}
where the factor $N_{sites}$ comes from summation over all $N_{sites}$ possible locations of $r_{h}$, and $\hat X \ket{\xi} = X\p{\xi} \ket{\xi}$. 

To perform the calculation we use a high-temperature expansion, $e^{-\gb \hat H} = \sum_{n=0}^{\infty} \frac{1}{n!}\p{-\gb \hat H}^{n}$. Each power of $H$ corresponds to a single ``hop'' of the hole, and the moments can be calculated from the sum of all closed paths of length $n$ (``$n-paths$'') starting at the origin,
\begin{equation} \begin{split}
\bra{\xi} \p{-\gb\hat H}^{n}\ket{\xi}  = \p{\gb t}^{n}\sum_{p\in n-paths}\braket{\xi}{\mathcal P_{p}\p{\xi}}.
\end{split} \end{equation}
Here $\mathcal P_{p}\p{\xi}$ is the spin permutation that results from moving the hole through the path $p$. Any open paths, that do not take the hole back to the origin, do not contribute to the sum, and the expectation value is zero if the path leads to a non-equivalent spin configuration. This requirement also restricts the sum to even values of $n$.

Although the number of closed paths grows exponentially with $n$, we are able to exhaustively enumerate them for small $n \le 2M$, and calculate a high temperature approximant
\begin{equation} \begin{split}
\avg{\hat X} & \approx \frac{N_{sites}}{Z}\sum_{n=0}^{M}\frac{\p{\gb t}^{2n}}{\p{2n}!} 
\sum_{p\in 2n-paths} \sum_{\xi}  X\p{\xi} \gd\p{\xi={\mathcal P}_p\p{\xi}}
\\ Z & \approx N_{sites} \sum_{n=0}^{M}\frac{\p{\gb t}^{2n}}{\p{2n}!} 
\sum_{p\in 2n-paths}  \sum_{\xi} \gd\p{\xi={\mathcal P}_p\p{\xi}}.
\label{eq:HTexp}
\end{split} \end{equation}
We use $M=6$.

Estimating the error of cutting off such series to be on the order of the last term calculated, the correlation functions for spins around the hole are accurate to about 10\% down to $T/t\sim 0.4$ for $M = 6$.  To investigate lower temperatures, one would need to resort to more sophisticated methods of summing the series, such as the Monte-Carlo approach of Raghavan and Elser \cite{Raghavan1995}.  Lower temperatures are difficult to achieve experimentally.

\section{Vacancy-Induced Ferromagnetism}

The tendency towards ferromagnetism is apparent in the structure of Eq.~(\ref{eq:HTexp}).  Ferromagnetic configurations $\xi$ automatically have  $\mathcal P\p{\xi}=\xi$, regardless of the path $p$.  A further insight is that it is only paths with loops in them that favor ferromagnetism.  Paths $p$ which retrace themselves have $\mathcal P\p{\xi}=\xi$ regardless of $\xi$.

\begin{figure}[ht]
\centering
\subfigure[$T\to\infty$]{
   \includegraphics[width=0.45\columnwidth] {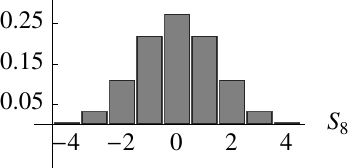}
 }
 \subfigure[$T\to 0$]{
   \includegraphics[width=0.45\columnwidth] {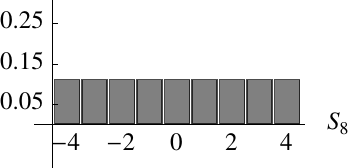}
 }
\caption{The probability distribution of $\avg{S_{8}}$, the total spin of the bosons around the hole, at high and low temperature.}
\label{fig:S8}
\end{figure}

To measure the polarization around the hole we define
\begin{equation} \begin{split}
\hat S_{8} = \sum_{i\in n.n.n}\hat S_{z}^{i}
\end{split} \end{equation}
where $S_{z}^{i}$ is the spin operator applied to the site $i$ and the summation is over the eight nearest-neighbor and next-nearest neighbor sites of the hole. 
The ground state of our system possesses a spontaneously broken symmetry. In an infinite system with an infinitesimal magnetic field along $z$, $\hat S_{8}$ will have a finite expectation value. This expectation value vanishes as $T\to\infty$ and approaches 4 as $T\to 0$. 
If there is no symmetry breaking field, then the spontaneous symmetry breaking occurs in a random direction.
In a typical cold-atoms experiment, every time a new sample is created, this symmetry-breaking direction will be different. Under those circumstances, one can model the ensemble measurement by taking expectation values in zero field. By symmetry, in zero field $\avg{\hat S_{8}}=0$, at all $T$, but the temperature dependence of its distribution will be non-trivial.
At $T\to\infty$ when all states are equally likely we expect a binomial distribution around zero. At $T\to 0$, 
the distribution is uniform. This may be understood in several ways; in a quantum mechanical treatment, one would attribute this to the fact that each projection $m$ of the spin multiplet is equally likely. Classically the z-component of a uniformly distributed random 3D unit vector is uniformly distributed. These distributions are shown in Fig.~\ref{fig:S8}.
 
To quantify these distributions, we examine the variance of $\hat S_{8}$. We define
\begin{equation} \begin{split}
\hat C_{8} & = \frac{3}{14}\br{\p{\hat S_{8}}^{2} - 2},
\\ & = \frac{3}{14}\br{\p{\sum_{i\in n.n.n.}\hat S_{z}^{i}}^{2} - \sum_{i\in n.n.n.}\p{\hat S_{z}^{i}}^{2}},
\end{split} \end{equation}
which is normalized and offset so that $\avg{C_{8}}$ goes to unity when the hole is maximally polarized and to zero when all sites are uncorrelated. Note that individual measurements of $\hat C_{8}$ can be negative or greater than one.

\begin{figure}[htbp] 
   \centering
   \subfigure[$\avg{C_{8}}$]{
	   \includegraphics[width=\columnwidth]{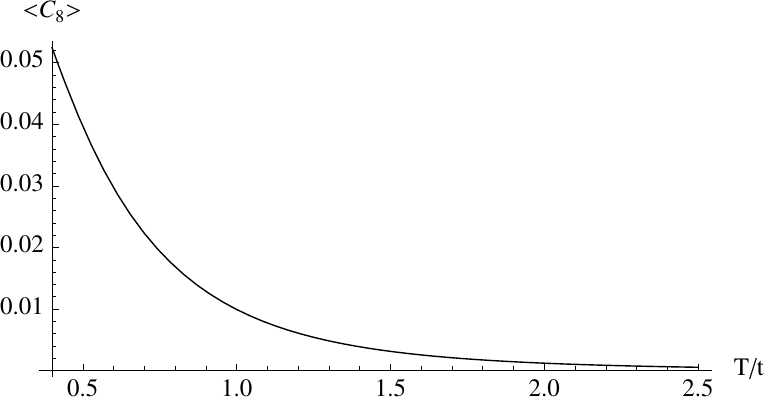} 
   }
   \\
   \subfigure[$\Delta C_{8}$]{
	   \includegraphics[width=\columnwidth]{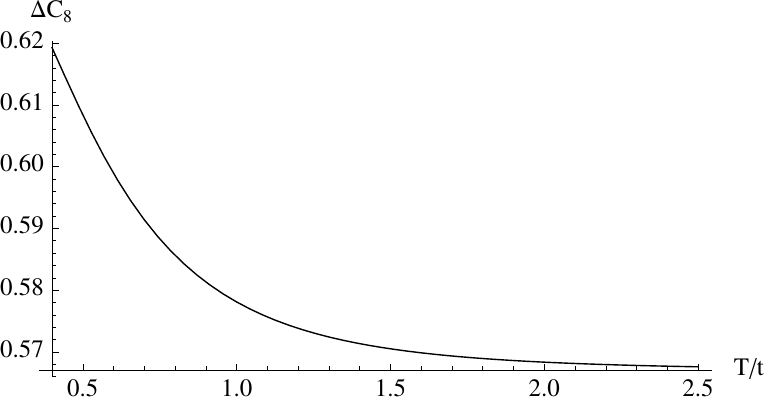} 
   }
   \caption{$\avg{C_{8}}$, the measure of polarization around the hole, and $\Delta C_{8} = \sqrt{\avg{{C_{8}}^{2}}-\avg{C_{8}}}$, as a function of the relative temperature $T/t$. Note that $\Delta C_{8}$ goes to $\sqrt{9/28}$ as $T\to\infty$.}
   \label{fig:C8}
\end{figure}

We have calculated for a range of temperatures  $\avg{C_{8}}$ and the uncertainty $\Delta C_{8} = \sqrt{\avg{{C_{8}}^{2}}-\avg{C_{8}}^{2}}$ and they are shown in Fig.~\ref{fig:C8}. In particular, at temperatures corresponding to $T/t = 0.4$ we predict $\avg{C_{8}} = 0.05$ and $\Delta C_{8} = 0.62$. This compares with a $T\to\infty$ result of $\avg{C_{8}} = 0$ and $\Delta C_{8} = \sqrt{9/28} \approx  0.57$. Both the non-zero mean of this quantity and the increase in variance are indicative of the ferromagnetic correlations present around the hole.  About $5000$ measurements would be needed  to determine the mean to within 20\% of the predicted value. A given sample will contain multiple holes, so each experimental run can contribute multiple independent measurements.

\section{Fixed Magnetization}

In a cold atom experiment the number of $\up$-spin and $\dn$-spin atoms are fixed, requiring a slightly different ensemble. This difference only matters when the correlation length becomes of the same order as the system size. For the temperatures described in Fig.~\ref{fig:C8}, the correlation length is of order the lattice spacing, and these subtleties are irrelevant.

By using exact diagonalization on a small system we can, however, show that at an order of magnitude lower temperature one must consider these finite size effects. We consider a system of $5\times 3$ sites described by the Hamiltonian in Eq.~\ref{eq:BH} with periodic boundary conditions, 7 $\up$-spins, 7 $\dn$-spins and a single hole. We define a similar operator to the one used before
\begin{equation} \begin{split}
\hat C_{8}^{f}  = \frac{3}{14}\br{\p{\hat S_{8}}^{2} - 2 - 56 C_{2}^{\infty}},
\end{split} \end{equation}
The constant $C_{2}^{\infty} = \avg{S_{z}^{1}S_{z}^{2}}$  is the infinite temperature two-spin correlation caused by the finite number of spins: $C_{2}^{\infty} = \frac{1}{4}\frac{1}{2-N_{\rm sites}}=-\frac{1}{52}$ for an equal number of $\up$ and $\dn$-spins.

\begin{figure}[htbp] 
   \centering
   \includegraphics[width=\columnwidth]{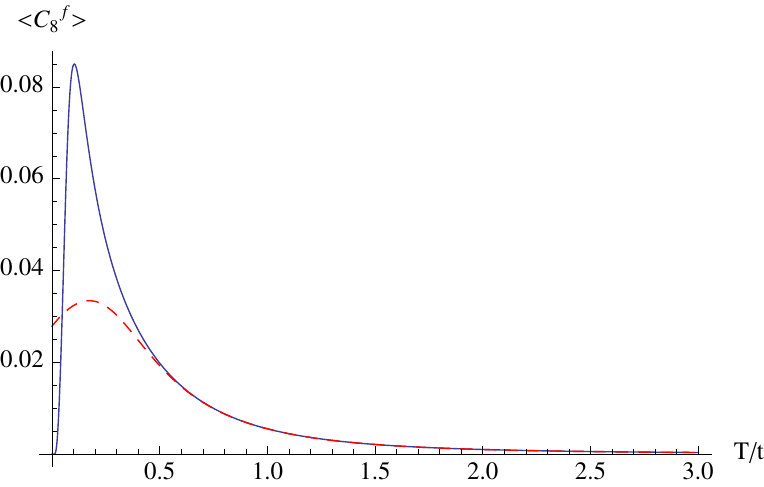} 
   \caption{(Color online) $\avg{\hat C_{8}^{f}}$ for a system of $3\times 5$ sites with an equal number of $\up$ and $\dn$ particles, as a function of temperature $T/t$. (solid blue line) Results from exact diagonalization, (dashed red line) result from high-temperature expansion taken to the same order as in Fig.~\ref{fig:C8}. The two match well to about $T/t\sim 0.4$.}
   \label{fig:threefive}
\end{figure}

The results are shown in Fig.~\ref{fig:threefive}. At high temperatures, one sees behavior indistinguishable from Fig.~\ref{fig:C8}, while at low temperatures the expectation value is suppressed. This suppression can be attributed to the ferromagnetic order parameter being forced to lie in the $x-y$ plane.

\section{Outlook}

The problem of how charge and spin degrees of freedom interact with one another is key to a number of important condensed matter systems, most notably high temperature superconductors. More importantly, conceptually clean examples of strongly correlated phenomena, such as the two component Bose system one, are essential to developing new paradigms for many-body physics.

In a cold gas experiment the quantities $\avg{S_8}$ and $\avg{C_8}$ can be measured by a variant of the quantum gas microscope technique pioneered by Bakr et al. \cite{Bakr2009} and extended to spinor gases by Fukuhara et al \cite{Fukuhara2012}. An image is taken of the optical lattice, which shows the location of all particles, and their spin projection along a fixed axis.  One would locate an isolated hole in this picture, and add up the spin projections of its neighbors to produce a single realization of $S_8$ or $C_8$.  The experiment would be repeated many times.  A histogram similar to Fig.~\ref{fig:S8} can be produced for $S_8$.  The ensemble average can be compared with our prediction for the quantum mechanical expectation value $\avg{C_8}$.

While the single-hole problem studied here is already interesting, the many-hole problem is even more rich. At zero temperature, the system is both superfluid and ferromagnetic. Kuklov et al. \cite{Kuklov2004,Kuklov2003,Kaurov2005}  have used Monte-Carlo methods to explore the relative strengths of superfluid and magnetic stiffnesses. Although no finite temperature studies have been done, both orders will disappear as one heats the system. It would be interesting to know if magnetism or superfluidity vanish first, or if the two orders vanish simultaneously \cite{Natu2011}. This question could be largely answered by studying the interaction between two polarons.

\section*{Acknowledgements}

This work was supported under by a grant from the ARO with funds from the  DARPA Optical Lattice Emulator program.

\bibliography{/Users/yarivyanay/Documents/University/Citations/library}
\bibliographystyle{apsrev}

\end{document}